\documentclass[aps,amsfonts,prl,nofootinbib,tightenlines,preprint]{revtex4}

\usepackage{bm}
\usepackage{epsfig}

\begin{document}

\title{Unnatural Oscillon Lifetimes in an Expanding Background}

\author{N.~Graham}
\email{ngraham@middlebury.edu}
\affiliation{Department of Physics, Middlebury College,
Middlebury, VT  05753}

\author{N.~Stamatopoulos}
\email{nstamato@middlebury.edu}
\affiliation{Department of Physics, Middlebury College,
Middlebury, VT  05753}

\preprint{\rm hep-th/0604134}

\begin{abstract}

We consider a classical toy model of a massive scalar field in $1+1$
dimensions with a constant exponential expansion rate of space.  The
nonlinear theory under consideration supports approximate oscillon
solutions, but they eventually decay due to their coupling to the expanding
background.  Although all the parameters of the theory and the oscillon
energies are of order one in units of the scalar field mass $m$, the
oscillon lifetime is exponentially large in these natural units.  For
typical values of the parameters, we see oscillon lifetimes scaling
approximately as $\tau \propto \exp(k E/m)/m$ where $E$ is the oscillon
energy and the constant $k$ is on the order of $5$ to $15$ for expansion
rates between $H=0.02m$ and $H=0.01m$.
 
\end{abstract}

\maketitle

\section{Introduction}

Central to many unsolved problems in particle physics and cosmology,
ranging from the flatness of the slow-roll inflaton potential to the
hierarchy of scales in grand unified theories, is the need to
understand the origin of ``unnaturally'' large or small dimensionless
parameters.  In this Letter we demonstrate how oscillons ---
localized, oscillatory solutions to nonlinear field equations that
either are stable or decay only after many cycles --- can provide an
explicit example in which such unnatural behavior emerges dynamically.
We show that in the presence of an expanding background metric,
oscillons in a simple toy model have lifetimes that are exponentially
large compared to the natural scales of the system, even though all
the inputs to the theory are of order one in these units.  Although
the situation is reminiscent of tunneling behavior in quantum systems,
this decay is entirely classical.

A wide variety of nonlinear field theories have been found to support
oscillon solutions (also known as breathers)
\cite{DHN,ColemanQ,Campbell,Bogolubsky,Gleiser,Honda,
iball,Wojtek,abelianhiggs,Forgacs,oscillon}.  In some special cases,
such as the sine-Gordon breather \cite{DHN} and $Q$-ball
\cite{ColemanQ}, conserved charges guarantee the existence of exact,
periodic solutions.  In $\phi^4$ theory in three dimensions, in contrast,
\cite{Bogolubsky,Gleiser}, oscillons are observed to
decay suddenly after lifetimes of order $10^3$ to $10^4$ in natural
units.  In other three-dimensional models \cite{oscillon}, no sign of
decay is visible in current numerical simulations.

In a broad class of one-dimensional models, perturbative analyses
\cite{DHN} and numerical simulations \cite{Campbell,abelianhiggs}
suggest that the oscillon lifetime is infinite, but analytic arguments
\cite{Kruskal} point to the existence of non-perturbative, exponentially
suppressed decay modes.  Such models then have a curious property:  in a
simple, classical theory in which all parameters are of order one in
natural units, classical dynamics generate a quantity --- the oscillon
lifetime --- that is exponentially large compared to the natural
scales of the system.

\section{Model}

In this Letter, we investigate explicitly a simple example of such a
system.  We consider a toy model consisting of a classical real scalar
field $\phi(x,t)$ of mass $m$ in one space dimension, governed by the
Lagrangian density
\begin{equation}
{\cal L} = \frac{1}{2}\left[(\partial_\mu \phi)^2 - m^2 \left(\phi^2
- \frac{1}{2} \phi^4 + \frac{1}{3} \phi^6\right)\right] \,.
\label{lagden}
\end{equation}
In this classical theory, the dynamics are unchanged when
the Lagrangian density is multiplied by a constant (equivalent to rescaling
$\hbar$). We have used this freedom, combined with a rescaling of the field
$\phi$, to fix the magnitude of the coefficient of the $\phi^4$ term 
while maintaining the conventional form of the free term.  The
arbitrary choice of the ratio of magnitudes of the $\phi^4$ and
$\phi^6$ terms is chosen for convenience and is not essential to our
results.  In the natural units of the mass $m$, then, all the
coefficients of the potential are of order $1$.  The only important
choices are the signs: the sign of the $\phi^2$ term gives the $\phi$
field a conventional (non-tachyonic) mass; the sign of the $\phi^4$
term is necessary for oscillons to exist; and the sign of the $\phi^6$
term is necessary for the field to be stable against runaway
growth.  We note that we have observed the same behavior in $\phi^4$
theory with the standard symmetry-breaking potential.  Here we have chosen
the $\phi^6$ model just to avoid any distractions that might be caused
by the existence of static ``kink'' solutions in the $\phi^4$ theory;
our theory contains no static solutions.

Like the other one-dimensional models discussed above, in a static
background the model of Eq.~(\ref{lagden}) supports oscillons that
appear to live indefinitely in numerical
simulations.  They are spatially localized, with size of order $1/m$
and fundamental frequencies of oscillation comparable to but always
below $m$.  We would like to include an additional element:  coupling
to an expanding background metric, inspired by the consideration of
oscillons in the early universe (work in progress \cite{group}; see
also studies of nontopological solitons and oscillons in hybrid
inflaton \cite{McDonald,Gleiserinflat} and axion-oscillons
\cite{Kolb}, and for a broader review of lattice field theory
simulations in the early universe see \cite{Smit}).  In comoving
coordinates, then, the Lagrangian takes the form
\begin{equation}
L = \frac{1}{2} \int \left[\dot \phi^2 - 
\frac{1}{a(t)^2} (\phi')^2 - m^2 \left(\phi^2
- \frac{1}{2} \phi^4 + \frac{1}{3} \phi^6 \right) \right] a(t) dx
\label{lag}
\end{equation}
leading to the equation of motion
\begin{equation}
\ddot \phi + \frac{\dot a(t)}{a(t)} \dot \phi = 
\frac{\phi''}{a(t)^2} - \phi + \phi^3 - \phi^5 \,.
\end{equation}
where the physical coordinate is related to the comoving coordinate
$x$ by the scale factor $a(t)$ and we hold the Hubble constant $H =
\dot a(t)/a(t)$ fixed, giving an exponential expansion rate.
Here $\dot \phi$ is the derivative of $\phi$ with respect to time 
and $\phi'$ is the derivative of $\phi$ with respect to the comoving
coordinate $x$.  This expansion destabilizes oscillons whose width is
of order of the horizon size $1/H$, but in numerical simulations we
observe that oscillons of smaller size remain stable, maintaining a
fixed size in physical units.

A technical advantage of this model is we can study oscillon behavior
efficiently for extremely long times, because any regions at
distances significantly greater than the horizon length $1/H$ from an
oscillon cannot influence its evolution.  Thus we do not need to
expand the box proportionally to the runtime \cite{oscillon} or
introduce absorbing boundaries \cite{Gleiserdamp} in order to prevent
unwanted reflections from disturbing the oscillon solution under study.

\section{Numerical Simulation}

Starting from thermal initial conditions with $T \gtrsim  m$, 
we see that oscillons emerge copiously as the expansion cools our
universe.  These initial conditions are generated using a canonical
ensemble of quantized modes of the free scalar field, and thus contain
no fine-tuning.\footnote{Here we must do a quantum calculation to
avoid the Jeans paradox.  As a result, $\hbar$ appears explicitly and
we can no longer scale out the overall normalization of the coupling
constants through our choice of units, as we have in the classical
calculation above.  The equal-time commutation relation $[
\phi(x,t), \dot \phi(y,t)] = i\hbar\delta(x-y)$ implies that $\phi$ has
units of $\sqrt{\hbar}$.  Therefore the Lagrangian density takes the form
\begin{equation}
L = \frac{1}{2} \int \left[\dot \phi^2 - 
\frac{1}{a(t)^2} (\phi')^2 - m^2 \left(\phi^2
- \frac{g}{2} \phi^4 + \frac{g^2}{3} \phi^6 \right) \right]
a(t) dx
\end{equation}
where $g$ is a constant with units of $1/\hbar$.  In the classical
theory, we may choose our units so this quantity is equal to $1$.  In
the thermal calculation, however, we encounter the quantity $\hbar
\omega/T$ in the Boltzmann factor for modes of energy $\omega$. Choosing
units so that $\hbar=1$ in this expression forces us to include $g$
explicitly in the classical Lagrangian density. An oscillon solution with
$g=1$ will continue to be a solution for general $g$, provided that we
scale up its amplitude by a factor of $1/\sqrt{\hbar g}$.  For our
classical treatment of the time evolution to be valid, the oscillon must
have an energy that is large compared to the energy of a typical quantum
fluctuation $\hbar \omega_0$, where $\omega_0 \approx m/\hbar$ is its
fundamental frequency of oscillation.  Since the oscillon energy goes like
$m/(\hbar g)$, this requirement can be implemented by choosing $\hbar g$ to
be small --- the classical approximation is valid at small coupling.  We
have verified that our results are not affected by the value of this
parameter; as long as $T \gtrsim  m/(\hbar g)$, oscillons are generated
copiously.  They then evolve according to the same classical dynamics
with the rescaled field.}  To simplify the classical analysis, we
neglect zero-point fluctuations.  Oscillons have been similarly found
to emerge spontaneously during phase transitions
\cite{Gleiserinflat,Gleiserphase}.

We use a straightforward numerical simulation in which we discretize
$x$ at the level of the Lagrangian in Eq.~(\ref{lag}), working in
natural units where $m=1$.  For the space derivatives we use ordinary
first-order differences,
\begin{equation}
\phi'(x_n,t) = \frac{\phi_{n+1}(t) - \phi_{n}(t)}{\Delta x}
\end{equation}
where $\phi_n$ refers to the value of $\phi$ at lattice point $n$.
We work on a regular lattice with spacing $\Delta x = 0.01$, and impose
periodic boundary conditions.  Varying this Lagrangian yields lattice
equations of motion with second-order space derivatives,
\begin{equation}
\ddot \phi_n(t) + \frac{\dot a(t)}{a(t)} \dot \phi_n(t) = 
\frac{\phi_{n+1}(t) + \phi_{n-1}(t) - 2 \phi_{n}(t)}{a(t)^2 \Delta x^2}
- \phi_n(t) + \phi_n(t)^3 - \phi_n(t)^5 \,.
\end{equation}
Finally, we use second-order differences 
\begin{equation}
\ddot \phi_n(t) = \frac{\phi_n(t+\Delta t) + \phi_n(t-\Delta t) -
2\phi_n(t)}{\Delta t^2}
\hbox{\quad and \quad}
\dot \phi_n(t) = \frac{\phi_n(t+\Delta t) - \phi_n(t-\Delta t)}{2\Delta t}
\end{equation}
with $\Delta t = 0.005$ to compute the field at $t+\Delta t$
based on the values at $t$ and $t-\Delta t$.  We have verified that
our results are not sensitive to the particular choice of lattice
spacing and time step.  (By the Courant condition we must choose
$\Delta t < \Delta x$ or the algorithm will be unstable.)

In the absence of expansion, our system conserves energy
exactly in the limit of infinitesimal time step $\Delta t$
(regardless of the spatial lattice size).  In an expanding background,
the configuration loses energy at a rate given by the pressure times
the rate of change in volume
\begin{equation}
\frac{dE}{dt} = -\int p(x,t) \dot a(t) dx
\end{equation}
where the pressure density is 
\begin{equation}
p(x,t) = \frac{1}{2}\left[\dot \phi^2 + 
\frac{1}{a(t)^2}(\phi')^2 - m^2 \left(\phi^2
- \frac{1}{2} \phi^4 + \frac{1}{3} \phi^6\right)\right] \,.
\end{equation}
We use this result to check the accuracy of our calculation, and find
that it is maintained to better than $1$ part in $10^4$, as we'd
expect for a second-order method.  As the universe expands, our
lattice expands with it, but the oscillons do not.  Therefore,
whenever our lattice has expanded by a factor of $2$, we refine the
lattice, doubling the total number of lattice points and bringing the
lattice spacing back to its original value in physical units.  We
assign values to the field for the new intermediate lattice points by
polynomial interpolation.  This numerical algorithm is highly stable,
maintaining precision without any sign of degradation even after
extremely long runs.

We begin our simulation using values of $\phi_n(t=0)$ and $\dot
\phi_n(t=0)$ drawn from a random thermal configuration with $T=1$
(setting the quantum scaling parameter $g$ to one for simplicity).  As
the expanding universe redshifts away the ordinary perturbative
fluctuations, it becomes simple to pick out the oscillon peaks.  
At this point, we excise a particular oscillon, keeping only a window
of size $100 + 2/H$ in physical units around it.  (We identify the
starting and ending points of the oscillon profile as the places where
the derivative of the energy density is $0.001$ in natural units.)  
Now, each time we insert new lattice points, we also truncate the
lattice back down to this size, so that the total computational cost
of the run scales only linearly with time.  Because of the expansion,
any noise introduced by this truncation can never affect the oscillon
dynamics.  We have verified that changing the box size used for this
truncation does not affect the oscillon lifetimes we observe.

\section{Decay Analysis}

Although the oscillon profiles we obtain by this method vary, they
all exhibit common behavior.  We use a small-amplitude asymptotic
analysis in a static background, following \cite{DHN,smallamp}, to gain
some insight into the general properties of oscillon solutions.  At
large $|x|$, the magnitude of $\phi$ is small and we can ignore the
nonlinear terms in the equations of motion, so our solution must be of
the form $\exp(-m|x|\epsilon) \sin \omega t$, where $\omega =
m\sqrt{1-\epsilon^2}$.  (We have absorbed an arbitrary phase by our
choice of the zero of $t$.)  Therefore it makes sense to work in terms
of rescaled variables $\chi = m x \epsilon$ and $\tau =
m t\sqrt{1-\epsilon^2}$, giving the equation of motion
\begin{equation}
\left( \phi \right)_{\tau \tau} + \phi = \epsilon^2 \left[
\left( \phi \right)_{\tau \tau} + \left( \phi \right)_{\chi\chi}
\right] + \phi ^3 - \phi^5 \,.
\end{equation}

Next we expand $\phi$ in powers of $\epsilon$ using these variables,
\begin{equation}
\phi(\chi, \tau) = \epsilon \phi_1(\chi, \tau) + 
\epsilon^2 \phi_2(\chi, \tau)  + \epsilon^3 \phi_3(\chi, \tau) + \ldots
\end{equation}
and consider the equations of motion order by order in $\epsilon$.  At
${\cal O}(\epsilon)$, we have simply
\begin{equation}
\left( \phi_1 \right)_{\tau \tau} + \phi_1 = 0
\end{equation}
and thus $\phi_1 = f(\chi) \sin \tau$, where we have again absorbed a
phase into the definition of $\tau$, and the profile $f(\chi)$ remains
to be determined.  At ${\cal O}(\epsilon^3)$, we have
\begin{equation}
\left( \phi_3 \right)_{\tau \tau} + \phi_3 = 
\left( \phi_1 \right)_{\tau \tau} + \left( \phi_1 \right)_{\chi\chi}
+ \phi_1^3 \,.
\end{equation}
The key point is that for all $\chi$, the function of $\tau$ on the
left-hand side of this equation is orthogonal to $\sin \tau$, so the
right-hand side must be as well.  As a result, setting the Fourier
coefficient of $\sin \tau$ on the right-hand side equal to zero gives
\begin{equation}
\frac{d^2 f}{d\chi^2} - f(\chi) + \frac{3}{4} f(\chi)^3 = 0
\end{equation}
whose solution, consistent with the boundary conditions at infinity, is 
$f(\chi) = \sqrt{\frac{8}{3}} \, {\rm \, sech \,} \chi$.  Thus we have
a family of approximate solutions 
\begin{equation}
\phi(x,t) \approx \epsilon \sqrt{\frac{8}{3}} \, 
\sin \left( m t \sqrt{1-\epsilon^2}\right)
{\rm \, sech \,} m x \epsilon
\end{equation}
parametrized by the small amplitude $\epsilon$.

\begin{figure}[htbp]
\includegraphics[width=0.5\linewidth,angle=270]{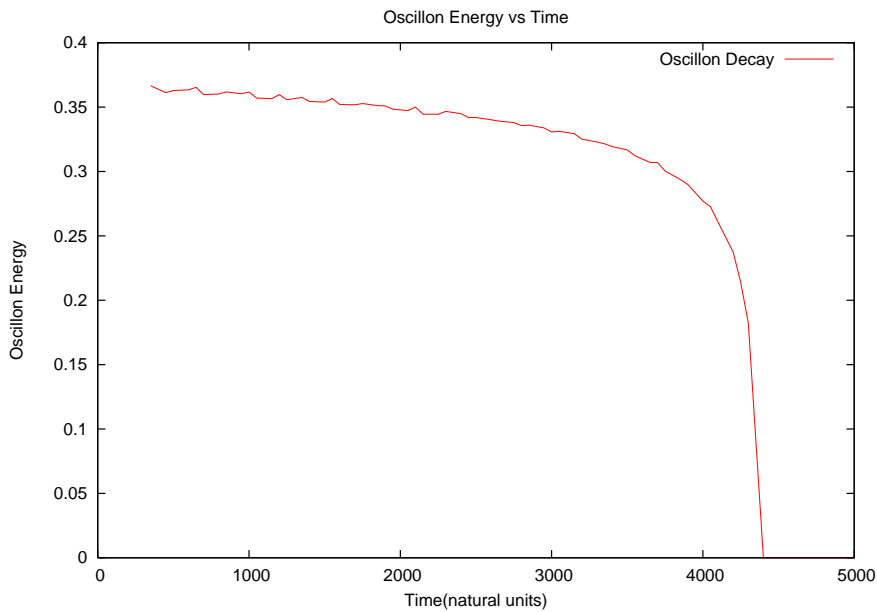}
\caption{
The final decay of an oscillon as it approaches the point of
instability.  Although the decay is very sudden, the approach to this
decay at earlier times is extremely gradual.  Here the energy is
measured in units of $m$ and the time is measured in units of $1/m$.}
\label{decay}
\end{figure}

Although this analysis is modified by higher orders in $\epsilon$ and
the inclusion of the expanding background, we find numerically that
its general features remain.  In particular, we can have oscillons of
large width as long as they have correspondingly small amplitudes; the
oscillon's total energy, which is proportional to its width times its
amplitude squared, scales linearly with its amplitude and inversely
with its width.  Each oscillon decays by a gradual process of
expansion, through which its width increases and its amplitude and
energy decrease.  As the width approaches the horizon size $1/H$, this
decay becomes very rapid and the oscillon decays suddenly.  This
process is shown in Fig.~\ref{decay}.  We define the decay of the
oscillon as the time when the derivative of its energy is everywhere
below $0.001$ in natural units.  Since the decay process is so sudden,
choosing any other reasonable criterion would make a negligible change
in the lifetime we measure.

\begin{figure}[htbp]
\begin{minipage}[h]{0.725\linewidth}
\includegraphics[width=0.7\linewidth,angle=270]{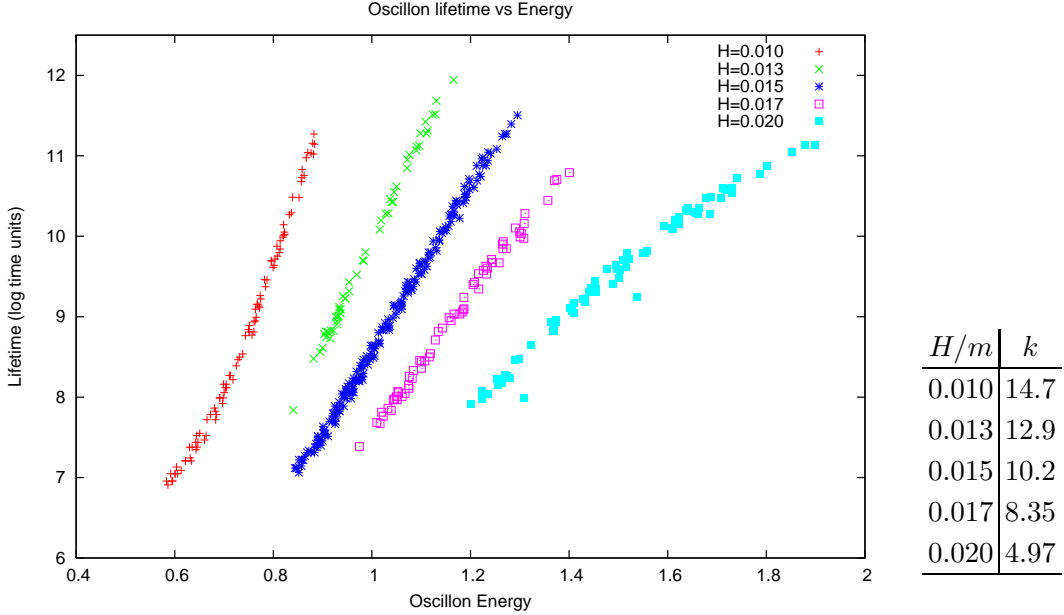}
\caption{Logarithmic plot of oscillon lifetime as function of the initial
oscillon energy for various values of the Hubble constant $H$.  The
vertical axis is the natural logarithm of the lifetime of the
oscillon and the horizontal axis is the oscillon energy, both in units
where $m=1$.  We see that the oscillon lifetime scales like $\exp(k
E/m)/m$, where $E$ is the oscillon energy.  The rate of exponential
growth $k$ is given in as a function of $H$ in the table at right.
The values of $k$ are roughly linear in $H$, with a slope of
approximately $-994/m$.}
\label{lifetimevsenergy}
\end{minipage}
\begin{minipage}[h]{0.15\linewidth}
\vfill
\begin{tabular}{c|c}
$H/m$ & $k$ \cr \hline 0.010 & 14.7\cr 0.013 & 12.9\cr 0.015 & 10.2\cr
0.017 & 8.35\cr 0.020 & 4.97\cr \hline
\end{tabular}
\vspace*{0.1in}
\vfill
\end{minipage}
\end{figure}

The dramatic behavior shown in Fig.~\ref{decay} is visible only at
the end of the oscillon's life, however.  For slightly higher
energies, this curve becomes extremely close to horizontal; oscillons
with only marginally higher energy lie far to the left.  As a result,
the oscillon lifetime scales exponentially with its energy, as shown
in Fig.~\ref{lifetimevsenergy}.  Here we have used the energy to
select oscillons that we predict will decay in a reasonable amount of
time.  However, we have also verified that more energetic oscillons
live for the exponentially long times that this analysis would
suggest.  In particular, we have followed an individual oscillon for
times exceeding $2 \times 10^7$ in natural units.  This analysis also
verifies that the decay is not the result of accumulated numerical error.

To track the flow of energy during the oscillon's decay, we consider a
fiducial box with a radius of one Hubble length $1/H$.
The rate of change in the energy within the box is then
\begin{eqnarray}
\frac{dE_{\rm box}}{dt}&=& 
\left. \frac{\dot \phi \phi'}{a(t)} \right|_{-1/H}^{1/H}
-\int_{-1/H}^{1/H} p \, \dot a(t) dx
\end{eqnarray}
where the first term represents the flow of of any outgoing waves through
the boundary and the second represents the energy lost to gravitational
expansion.  For our oscillon configurations, both terms oscillate.  On
average, however, we find the surprising result that the first term is both
positive and much smaller in magnitude than the second term, meaning that
the oscillon decays due to energy lost through the expansion rather than
through the emission of scalar field waves.

Finally, we note that although the expansion of the universe
explicitly introduces an exponentially large scale --- the size of the
universe --- it is irrelevant to the oscillon lifetime.  The oscillon
only sees its local region of space, which is expanding at a rate of
order one in natural units, with greater expansion of the universe
corresponding to \emph{shorter} oscillon lifetimes.

\section{Conclusions}

We have found that oscillon lifetimes in an expanding background scale
exponentially with the oscillon's energy in a simple toy model.  The
oscillon configurations form generically from an uncorrelated thermal
background.  As a result, although all the parameters of our theory are of
order one in natural units, the dynamics of the theory generates an
exponentially large time scale through the oscillon's decay.  Although
our analysis is limited to one dimension, we have also conducted
preliminary experiments showing similar behavior when the
three-dimensional oscillons of Refs.~\cite{Gleiser,oscillon} are coupled 
to an expanding background in an ansatz with spherical symmetry.  Work
is in progress \cite{group} to explore the possible consequences of
these ideas in the early universe.

\section{Acknowledgments}

We thank N.~Alidoust, E.~Farhi, A.~H.~Guth, A. Scardicchio, R. Rosales, and
R.~Stowell for assistance.  We also thank J.~Dunham and M.~Gleiser for
discussions.  N.~G. and N.~S. were supported in part by Research Corporation
through a Cottrell College Science Award.  N.~G. was also supported in
part by the National Science Foundation through the Vermont
Experimental Program to Stimulate Competitive Research (VT-EPSCoR) and
the Middlebury Faculty Professional Development Fund.  N.~S. was also
supported in part by the Middlebury Undergraduate Collaborative
Research Fund.

\end{document}